\documentclass[twocolumn,pra,floatfix,showpacs]{revtex4}

\usepackage{graphicx}
\usepackage{dcolumn}
\usepackage{amsmath}
\usepackage{amssymb}

\begin{document} 

\title{Universal Quantum Logic from Zeeman and Anisotropic Exchange
Interactions}

\author{Lian-Ao Wu and Daniel A. Lidar}

\affiliation{Chemical Physics Theory Group, Chemistry Department, University of Toronto, 80 St. George Str., Toronto, Ontario M5S 3H6, Canada}

\begin{abstract}
Some of the most promising proposals for scalable solid-state quantum
computing, e.g., those using electron spins in quantum dots or donor
electron or nuclear spins in Si,
rely on a two-qubit quantum gate that is ideally generated by an
isotropic exchange interaction.  However, an anisotropic perturbation
arising from spin-orbit coupling is inevitably present. Previous
studies focused on removing the anisotropy. Here we introduce a new
universal set of quantum logic gates that {\em takes advantage} of the
anisotropic perturbation.  The price is a constant but modest factor
in additional pulses. The gain is a scheme that is compatible with the
naturally available interactions in spin-based solid-state quantum
computers.
\end{abstract}


\maketitle 

\section{Introduction}

A fundamental notion in quantum computing (QC) is {\em universality}: a set
of quantum logic gates (unitary transformations) is said to be
``universal for QC'' if any
unitary transformation can be approximated to arbitrary accuracy by a
quantum circuit involving only those gates \cite{Nielsen:book}. Mathematically,
this means the ability to efficiently generate a dense subgroup of the group of unitary
operations on $N$ qubits, $U(2^{N})$. Physically, this is accomplished by
carefully manipulating single qubit-external field and (or only) qubit-qubit
interactions, thus generating unitary gate operations. A {\em universal-gate
set} that accomplishes this, may be continuous, discrete, or both. A
well-known example is the set of all single-qubit gates plus a
controlled-phase ({\sc CP}) gate (that flips the phase of a target
qubit depending on the state of a control qubit), but many other universal sets are
known \cite{Nielsen:book}. An important example of a universal gate set, of
relevance to us, is the set generated by controlling only {\em isotropic}
Heisenberg exchange interactions. This set was shown \cite{Bacon:99aKempe:00} to be universal in the
context of research on decoherence-free subspaces (DFSs) \cite{Zanardi:97cDuan:98Lidar:PRL98}, and requires that a logical qubit be encoded into at
least $3$ physical qubits \cite{Knill:99a}. Efficient gate sequences
for universal QC in this case were subsequently presented in Ref.~\cite{DiVincenzo:00a}. These
results assume that all qubits have equal energies. However, this assumption
may break down under magnetic field and/or $g$-factor inhomogeneity  \cite{Hu:01a}. When
the resulting Zeeman splitting is taken into account, it can be shown that
the isotropic Heisenberg interaction is universal for QC using an encoding of one
logical qubit into only two physical qubits, and efficient gate sequences
have been found \cite{Levy:01Benjamin:01,LidarWu:01}. We describe here a new universal gate
set: that generated by the Zeeman splitting and the {\em anisotropic} Heisenberg
interaction, defined more precisely below. This set is of particular
importance to spin-based solid-state approaches to quantum computing
\cite{Loss:98Kane:98,Vrijen:00}, where anisotropy is inherently
present \cite{Kavokin:01}.

\section{Zeeman and Exchange Interactions}
A single spin $
\vec{S}=(S^{x},S^{y},S^{z})$ with magnetic moment $\mu _{B}$
couples to a magnetic field $B(t)$ oriented along the $z$ axis through the
Zeeman splitting Hamiltonian $g\mu _{B}B(t)S^{z}$. This interaction can be spatially
controlled by making $B(t)$ inhomogeneous \cite{Loss:98Kane:98}, or by
modulating the $g$-factor \cite{Vrijen:00}. Conversely, inhomogeneities
and/or a non-uniform $g$-factor may be naturally present \cite{Hu:01a}. The Zeeman
splitting removes the degeneracy of the two spin states and serves to
define a physical qubit. Switching on the Zeeman term for the $j^{{\rm th}}$
qubit causes a phase shift, i.e., it generates the single-qubit gate $
e^{-i\eta S_{j}^{z}}$, where 
\[
\eta =\int dt\,g\mu _{B}B(t)
\]
is a controllable parameter (we use units where $\hbar =1$). E.g., a useful
gate is $Z_{j}=i\exp (-i\pi S_{j}^{z})$, which is a 180$^{0}$ rotation about
the $z$ axis. The typical switching time of the Zeeman splitting is fast:
it is similar to
that of the Heisenberg interaction (GHz), which is the
interaction assumed to govern the operation of two-qubit gates in some of
the spin-based approaches to quantum computing \cite{Loss:98Kane:98,Vrijen:00}. These QC proposals, as well as schemes for
universal QC using the Heisenberg interaction alone \cite{Bacon:99aKempe:00,DiVincenzo:00a} rely on this interaction being perfectly
isotropic. However, in a crystal environment that lacks inversion symmetry, the actual
interaction between spins $i$ and $j$ is
\begin{equation}
H_{ij}(t)=J(t)(\vec{S}_{i}\cdot \vec{S}_{j}+\vec{\beta }(t)\cdot 
\vec{S}_{i}\times \vec{S}_{j}+\gamma(t) \vec{\beta }(t)\cdot \vec{S}_{i}
\vec{\beta }(t)\cdot \vec{S}_{j}),
\label{eq:Hij}
\end{equation}
where only the exchange parameter $J(t)$ is directly
controllable \cite{Kavokin:01}. This means that the isotropic Heisenberg interaction $J(t)\vec{
S}_{i}\cdot \vec{S}_{j}$ itself is {\em not} independently
tunable. The anisotropic part arises from spin-orbit coupling, as a
relativistic correction. 
As
written, the anisotropy parameters $\vec{\beta }$ and $\gamma $
are dimensionless; in systems like coupled GaAs quantum dots
$|\vec{\beta }|$ is of the order of a few
percent, while the last term is of the order of $10^{-4}$ \cite{Kavokin:01}. $H_{ij}(t)$ given in Eq.~(\ref{eq:Hij}) is the
most general anisotropic exchange interaction that is symmetric about a
\emph{given axis}, here $\vec{\beta }$. Further corrections will be
even smaller. The
anisotropic perturbation has been considered a problem and strategies have
been designed to cancel it. E.g., it can be removed to first order by shaped
pulses \cite{Bonesteel:01}, or cancelled in the absence of an external
magnetic field \cite{Burkard:01}. Instead of
trying to cancel the anisotropy, we
show here how to use it to our advantage in order to generate a universal
gate set.

We first focus on the case of time-{\em in}dependent $\vec{\beta }
$ and $\gamma $, which should be dominant as discussed recently in Ref.~\cite{Burkard:01}. The corrections arising from the time-dependent
anisotropic interaction are much weaker, sufficiently so that they are below
the threshold for fault tolerant quantum computation \cite{Burkard:01,Aharonov:99}. Nevertheless, we also consider the time-dependent
case below. Now, turning on the exchange term $H_{ij}(t)$ generates a
unitary evolution 
\[
U_{ij}(\varphi )=\exp (-i H_{ij}(\varphi ))
\]
through the Schr\"{o}dinger equation, where $H_{ij}(\varphi ) \propto
\varphi$ and
\[
\varphi =\int dt J(t)
\]
is a second controllable parameter.

We assume that {\em we can only use the two parameters} $\eta $ {\em and} $
\varphi $ to manipulate computational states and construct a universal gate
set. Direct control of Hamiltonian terms that generate single-qubit
rotations about the $x$ and $y$ axes causes device heating and other major
technical problems, so that this type of control is best avoided \cite{DiVincenzo:00a,LidarWu:01}. We thus refer to $H_{ij}$ and the Zeeman
splitting as the ``available Hamiltonians''. We now show that using control
only over these available Hamiltonians suffices to generate universal gate
sets for a variety of orientations of the vector $\vec{\beta }$.

\begin{figure}
\includegraphics[height=13cm]{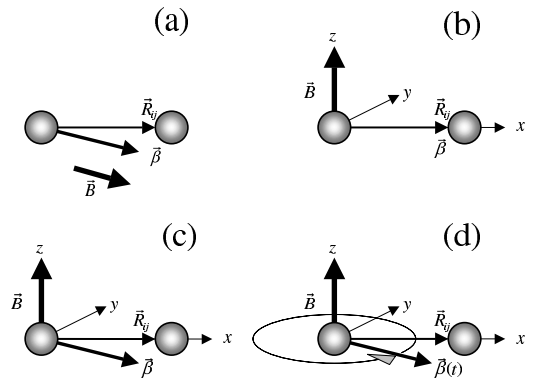}
\vspace{-7cm}
\caption{\label{fig1}Geometries of magnetic field $\vec{B}$, relative position of
quantum dots $\vec{R}_{ij}$, and spin-orbit field $\vec{\beta}$,
considered in the text. Quantum dots are indicated by shaded circles.}
\end{figure}

Following Ref.~\cite{Kavokin:01}, the orientation of
$\vec{\beta }$ is expressed in terms of the vector
$\vec{R}_{ij}$ pointing from qubit $i$ (e.g., the center of
the $i^{{\rm th}}$ quantum dot) to qubit $j$ (Fig.~\ref{fig1}). We can always
choose the direction of the magnetic field as the $z$ axis. Since
$\vec{R }_{ij}$ is a vector in the plane the quantum dots
are lying on, if the magnetic field is applied parallel to
$\vec{\beta }$, it too should be in the plane of the dots
(Fig.~\ref{fig1}a). A more common case is when the magnetic field is
perpendicular to the plane of the dots (Figs. 1b-d). We proceed to
analyze each of these four cases.

\section{Case 1: Magnetic Field Parallel to $\vec{\beta }$}
\begin{figure}
\includegraphics[height=13cm]{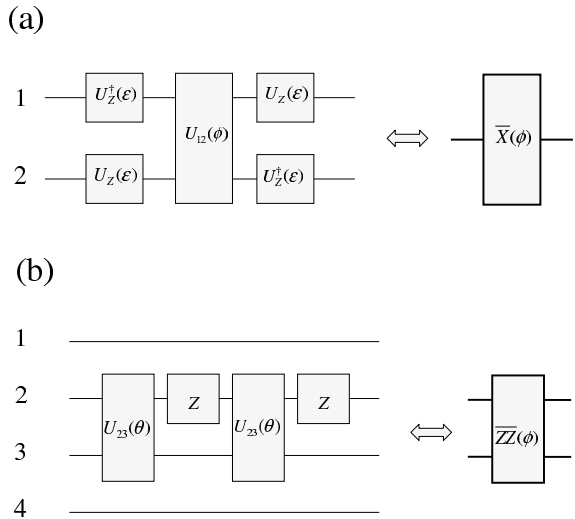}
\vspace{-5cm}
\caption{\label{fig2}Diagrams of circuits implementing logical
operations in the $\vec{\beta} || \vec{B}$ case. Lines denote physical
qubits, time flows from left to right. The $3$-step circuit (a)
implements the transformation $
\overline{X}(\phi) \equiv \exp (-i\phi \overline{S^x}_{1}) =
\exp (-i\epsilon \overline{S^z}_{1}) U_{12}(\phi/\sqrt{1+\beta^2})
\exp (i\epsilon \overline{S^z}_{1})$.
Here $\phi $ is an arbitrary angle and $\epsilon =\arctan\beta $ is
a time-independent constant
\protect\cite{comment2}. The $4$-step circuit (b)
implements the transformation $
\overline{ZZ}(\phi) \equiv \exp (i\phi
\overline{S^z}_{1}\overline{S^z}_{2}) = U_{23}(\theta
)Z_{2}U_{23}(\theta )Z_{2}$, where $\theta =\frac{1}{2}\phi /(1+\gamma
\beta ^{2})$. The notation used in the diagrams is: $U_Z(\epsilon)
\equiv \exp (-i\epsilon \overline{S^z})$, $Z \equiv i\exp (-i\pi S^{z})$.}
\end{figure}

We first
discuss the case in which the magnetic field $\vec{B}$ is parallel to $\vec{
\beta }$ ($=\beta $ $\vec{e}_{z}$, Fig.~\ref{fig1}a). In this (and
only this) case it was shown in 
Ref.~\cite{Burkard:01} that the effect of the anisotropy may be made to cancel exactly. 
However, this approach requires precise alignment of $\vec{B}$ along $\vec{
\beta }$, and utilizes single-qubit $S^x,S^y$ interactions
for universality, which as discussed above, we seek to avoid here.
Indeed, in the $\vec{\beta} || \vec{B}$ case, the available
Hamiltonians are not universal for QC because they have too much symmetry: $H_{ij}$
and the Zeeman splitting both commute with $S_1^{z}+S_2^z$. There is
a simple way to solve the problem: we encode a pair of physical
qubit states into a logical qubit:\ $|0_{L}\rangle =|\uparrow \rangle
|\downarrow \rangle $ and $|1_{L}\rangle =|\downarrow \rangle |\uparrow
\rangle $ (see Refs.~\cite{Levy:01Benjamin:01,LidarWu:01,WuLidar:01} for other cases where this
encoding proved useful for universality). In this manner the first logical
qubit is given by physical qubits $1,2$, the second by physical qubit $3,4$,
and so on. A calculation then shows that the {\em encoded} (denoted by a
bar) single-qubit operations are: $\overline{S^x}_{i}=\vec{S}_{2i-1}\cdot \vec{S}
_{2i}-S_{2i-1}^{z}S_{2i}^{z}$, $\overline{S^y}_{i}=-(\vec{S}_{2i-1}\times \vec{S}
_{2i})_{z}$, and $\overline{S^z}_{i}=(S_{2i-1}^{z}-S_{2i}^{z})/2$, where the
subscript denotes the $i^{\rm th}$ encoded qubit. These operators have the same
commutation relations as the three components of spin angular momentum
[i.e., they generate $su(2)$]. Under our assumption of a controllable Zeeman
splitting, we can switch $\overline{S^z}$ on/off, and hence can perform
arbitrary rotations about the $z$ axis of the encoded qubit. While we do not
have direct access to $\overline{S^x}$, the $3$-step quantum
circuit depicted in Fig.~\ref{fig2}a yields this operation.

The ability to perform arbitrary rotations about the $z$ and $x$ axes
suffices for performing arbitrary single qubit operations, through a
standard Euler angle contruction \cite{Nielsen:book}. To complete the
universal gate set we also need a logic gate coupling different encoded
qubits in a non-trivial manner, such as a {\sc CP} gate. We have
previously shown that the interaction $\overline{S^z}_{1}\overline{S^z}_{2}$
between logical qubits $1$ and $2$, and which generates a {\sc CP}\ gate
between these qubits, is equivalent to the interaction $S_{2}^{z}S_{3}^{z}$
between physical qubits $2$ and $3$ \cite{WuLidar:01}. This can be
implemented by the $4$-step quantum circuit depicted in
Fig.~\ref{fig2}b. We note that this circuit also provides a way to cancel the anisotropic
interaction by controlling the Zeeman splitting (see also Ref.~\cite{Burkard:01}). In addition, the encoded qubit is a DFS against collective dephasing errors \cite{Bacon:99aKempe:00}, so an
automatic layer of error protection is built into these circuits.

\section{Case 2: Magnetic Field Perpendicular to Plane of Dots}
We now
analyze the more common case where the
magnetic field is perpendicular to the $x-y$ plane the quantum dots
are on. First we consider $\vec{\beta }=\beta \vec{e}_{x}$ (or $\beta \vec{e}_{y}$), which can be
along the direction $\vec{R}_{12}$ from qubit $1$ to qubit
$2$ (Fig.~\ref{fig1}b). 
As shown in the previous case (of Fig.~\ref{fig1}a), the isotropic Heisenberg interaction and Zeeman
splitting become universal for QC by using an encoding. In contrast, as
we now show, $H_{ij}$ together with the Zeeman splitting are
universal {\em without encoding}. Since $S^{z}$ is by our assumptions
controllable, the problem for single-qubit rotations is to show how to
generate $S^{x}$. We will explicitly be using the anisotropic perturbation
to this end, so the speed of the $S^{x}$ gate will be on the order of a few
percent of the $S^{z}$ gate. This is still reasonable since it is similar
to, or even better than, the relative strength of the two-spin interaction
and the external radiofrequency magnetic fields in NMR \cite{Nielsen:book}.

To generate the $S^{x}$ gate we first introduce a simple $3$-step quantum
circuit, that will serve as a building block for other gates: 
\[
V=U_{12}(\pi /\sqrt{1+\beta ^{2}})Z_{1}Z_{2}U_{12}(-\pi /\sqrt{1+\beta ^{2}}),
\]
(note that $Z_{1}Z_{2}$ can be implemented in one parallel step). Contrary to its appearance,
this gate is actually separable for qubits $1$ and $2$. This allows us to
use it for creating single-qubit gates, e.g., the following $8$-step
circuit: 
\[
\exp (-i4\epsilon S_{1}^{x})=Z_{1}VZ_{1}V.
\]
Recall that $\epsilon =\arctan\beta $ and $\beta $ is fixed (given) so
that this circuit is {\em discrete} (only whole multiples of the angle $
4\epsilon $ can be generated). Alternatively, the following $17$-step
circuit yields a continuous $S^{x}$ gate \cite{comment1}: 
\begin{eqnarray*}
\exp (i\phi S_{1}^{x})=\exp (i\delta S_{1}^{z})V\exp (i\eta
S_{1}^{z})V \times \\
Z_{1}Z_{2}V\exp (-i\eta S_{1}^{z})VZ_{1}Z_{2}\exp (-i\delta
S_{1}^{z})
\end{eqnarray*}
with $\delta =\pi /2-\arctan(\tan \eta /2\cos 2\epsilon )$, and
continuous angle $\phi =2\arccos (1-2\sin ^{2}\eta /2\sin ^{2}2\epsilon )$,
controlled in terms of $\eta $. However, $|\phi |$ is bounded because of $
\epsilon $. E.g., its maximum is approximately $\pi /12$ if $\epsilon =0.03.$
Therefore, in order to perform exactly a single-qubit $S^{x}$ gate with
larger angle, one can first use $V$ to approximate the needed gate and then
converge using the $17$-step continuous gate.

To complete the discussion of universality we again need to generate a logic
gate coupling qubits. Such a two-qubit operation can be obtained in terms of
the following (not necessarily optimized) $55$-step quantum circuit,

\begin{eqnarray*}
\exp (-i\phi S_{1}^{z}S_{2}^{z})=Z_{2}\exp (-i\frac{\epsilon }{2}
(S_{1x}-S_{2x}))U_{12}(\varphi )Z_{2} \times \\
\exp (i\epsilon S_2^{x})U_{12}(\varphi
)\exp (i\frac{\epsilon }{2}(S^x_{1}-S^x_{2})), 
\end{eqnarray*}
where the arbitrary angle $\phi =2\varphi \sqrt{1+\beta ^{2}}$ is controlled
in terms of the angle $\varphi $ in $H_{ij}$. This gate is therefore no
longer slow. Note that $e^{i\phi (S^x_{1}-S^x_{2})}$ can be implemented as
above by a $17$-step quantum circuit. Further note that since {\em any}
entangling gate is universal (together with single-qubit gates)
\cite{Dodd:01}, in practice one may be able to reduce our $55$-step
circuit, e.g., using geometric time-optimal control methods \cite{Khaneja:02}.

\section{Case 3: General Time-Independent Case}

The general $\vec{\beta }\bot \vec{B}$ case is where $
\vec{\beta }=\beta _{x}$ $\vec{e}_{x}+\beta _{y}$ $
\vec{e}_{y}$, i.e., time-independent and somewhere in the $x-y$
plane (Fig.~\ref{fig1}c). However, this case is equivalent up to a unitary
rotation to the $\vec{\beta} = \beta \vec{e}_x$
case. Specifically, the transformation $e^{i\omega
(S_{1}^{z}+S_{2}^{z})}U_{12}(\varphi ) e^{-i\omega
(S_{1}^{z}+S_{2}^{z})}$ (where $\omega
=\arctan(\beta _{y}/\beta _{x})$), rotates $\vec{\beta}$ so that it becomes parallel to
$\vec{e}_x$. The treatment above then applies provided we everywhere replace $\beta$ by $\sqrt{\beta _{x}^{2}+\beta
_{y}^{2}}$.

It is noteworthy
that in the present case of time-independent $\vec{\beta }$,
similarly to Ref.~\cite{DiVincenzo:00a} where efficient gate sequences for the
isotropic Heisenberg interaction were obtained, we did not employ the
short-time approximation, i.e., made use only of finite time steps. In
contrast to the numerically derived circuits of
Ref.~\cite{DiVincenzo:00a}, our 
circuits are based on analytical results, and can be understood using
elementary angular momentum theory \cite{comment1,comment2}.

\section{Case 4: General Time-Dependent Case}

Finally, we also consider the general case with $\vec{\beta }$
and $\gamma $ both time-dependent, $\vec{\beta }(t)$ in the $x-y$
plane (Fig.~\ref{fig1}d). In contrast to the time-independent case, gates now have to be implemented
using the short-time approximation: $e^{A\Delta
t}e^{B\Delta t} = e^{(A+B)\Delta t + O(\Delta t^2)}$ for operators $A$ and $B$ that do not necessarily commute,
and $\Delta t\ll 1$. While this is less accurate than the exact circuits
given above, it is nevertheless a valuable and common tool in discussions of
universality \cite{Nielsen:book,Bacon:99aKempe:00,Lloyd:95}. The short-time evolution
operator\ corresponding to $H_{ij}$ has the same form as before: $
U_{12}(\Delta \phi )$ with $\Delta \phi =J\Delta t$, except that now $J$ is
an average value of the coupling constant in the time interval from $0$ to $
\Delta t$. Assuming that all time-dependent parameters do not vary
appreciably within the short time $\Delta t$, a two-qubit {\sc CP} gate
is given by the repeated $4$-step circuit 
\begin{equation}
\exp (-i\phi S_{1}^{z}S_{2}^{z})\approx (U_{12}(\phi
/4n)Z_{1}Z_{2}U_{12}(\phi /4n)Z_{1})^{2n},
\label{eq:short-t-CPHASE}
\end{equation}
where $\phi =n\Delta \phi$. The approximation improves with increasing $n$. Since $\phi/4n \ll 1$ we only need to know the detailed properties of
the evolution operator around time zero. Next we must generate the
single-qubit $S^{x}$ gate. To do so we combine a short-time and a
finite-time circuit. First, 
\begin{eqnarray*}
e^{-i\phi (S_{1}^{z}S_{2}^{y}-S_{1}^{y}S_{2}^{z})} \approx (e^{i\omega
(S_{1}^{z}+S_{2}^{z})}U_{12}(-\Delta \phi )Z_{1}Z_{2} \times \\
U_{12}(\Delta
\phi )Z_{1}Z_{2} e^{-i\omega (S_{1}^{z}+S_{2}^{z})})^{n} 
\end{eqnarray*}
where $\phi =n\Delta \phi \sqrt{\beta _{x}^{2}+\beta _{y}^{2}}$ and we have
used the short-time approximation. Then the single-qubit $S^{x}$ gate is
given in terms of the following circuit: 
\begin{eqnarray*}
e^{-i\phi S_{1}^{x}} = e^{i{\pi }S_{1}^{z}S_{2}^{z}} 
e^{-i\phi (S_{1}^{z}S_{2}^{y}-S_{1}^{y}S_{2}^{z})} Z_{2} \times \\
e^{-i\phi
(S_{1}^{z}S_{2}^{y}-S_{1}^{y}S_{2}^{z})} e^{-i{\pi }
S_{1}^{z}S_{2}^{z}} Z_{2}.
\end{eqnarray*}
This completes the generation of single-qubit gates, and thus proves universality
of our available interactions in the time-dependent case.

\section{Managing decoherence}

A discussion of universal quantum
computation is incomplete without a consideration of decoherence, the
process whereby quantum information is degraded through the interaction of
qubits with their environment. In principle three of the major methods
for resisting decoherence, quantum error correcting codes
\cite{Nielsen:book,Knill:99a,Aharonov:99}, DFSs
\cite{Bacon:99aKempe:00,Zanardi:97cDuan:98Lidar:PRL98,Knill:99a}, and
fast/strong ``bang-bang'' (BB) pulses \cite{Viola:98Vitali:99} are compatible with our universality results. As mentioned above,
in the case of $\vec{\beta }||\vec{B}$ we have used an
encoding into a DFS that is automatically resistant to collective dephasing
errors. We have recently shown how, starting from a general (linear)
system-bath coupling, to actively create the conditions for collective
decoherence by applying BB pulses generated by the isotropic Heisenberg
interaction \cite{WuLidar:01b}. In this case an encoding into a $3$- or $4$-qubit DFS is possible, which resists the remaining collective errors. Leakage errors (which would
arise due to corrections to the short-time approximation invoked in
BB theory) can likewise be
eliminated using only the isotropic Heisenberg interaction \cite{Kempe:01WuByrdLidar:02}.
We conjecture that the same (creation of collective decoherence, leakage
elimination) should be possible using the available interactions we
considered here. Even without encoding, the use of BB pulses should serve to
significantly enhance the robustness of our circuits under decoherence.

We further note that some of our circuits already have a form of
decoherence-resistance built into them. E.g., the form of Eq.~(\ref{eq:short-t-CPHASE}) is that of a parity-kick operation \cite{Viola:98Vitali:99}, which implies that this circuit eliminates all
Hamiltonians (including system-bath) containing system operators which
anti-commute with $Z_{1}Z_{2}$ and $Z_{1}$ or $Z_{2}$. In fact the same
consideration shows that this circuit also eliminates the undesired
anisotropic interaction in the more complicated case in which the strength
of the anisotropic interaction is not proportional to that of the Heisenberg
interaction.

\section{Conclusions}

We have introduced a new set of universal
Hamiltonians: the Zeeman splitting and anistropic Heisenberg interaction. This set
is of direct relevance to quantum computing in solid state systems that rely
on spin-spin interactions \cite{Loss:98Kane:98,Vrijen:00}. Until recently,
most studies of such systems assumed an isotropic Heisenberg interaction,
which, however, is an approximation due to spin-orbit coupling and other
perturbations \cite{Hu:01a,Kavokin:01}. 
Instead of trying to cancel the resulting anisotropy
\cite{Bonesteel:01,Burkard:01}, we showed here how to advantageously use the anisotropy. We analytically
derived circuits which implement universal quantum logic in a variety of
geometries of interest, for both time-independent and time-dependent
perturbations. In the former case, depending on geometry and type of gate implemented,
these circuits come with an overhead of between $3$ and at most $55$ extra pulses. 
We hope that the methods presented here will enhance the
prospects of quantum information processing in those promising quantum
computing proposals where the inherent anisotropy of the exchange
interaction cannot be ignored.

\begin{acknowledgments}
This material is based on
research sponsored by the Defense Advanced Research Projects Agency under
the QuIST program and managed by the Air Force Research Laboratory (AFOSR),
under agreement F49620-01-1-0468 (to D.A.L.). The U.S. Government is authorized to
reproduce and distribute reprints for Governmental purposes notwithstanding
any copyright notation thereon. The views and conclusions contained herein
are those of the authors and should not be interpreted as necessarily
representing the official policies or endorsements, either expressed or
implied, of the Air Force Research Laboratory or the U.S. Government.
\end{acknowledgments}


\begin{thebibliography}{32}
\expandafter\ifx\csname natexlab\endcsname\relax\def\natexlab#1{#1}\fi
\expandafter\ifx\csname bibnamefont\endcsname\relax
  \def\bibnamefont#1{#1}\fi
\expandafter\ifx\csname bibfnamefont\endcsname\relax
  \def\bibfnamefont#1{#1}\fi
\expandafter\ifx\csname citenamefont\endcsname\relax
  \def\citenamefont#1{#1}\fi
\expandafter\ifx\csname url\endcsname\relax
  \def\url#1{\texttt{#1}}\fi
\expandafter\ifx\csname urlprefix\endcsname\relax\def\urlprefix{URL }\fi
\providecommand{\bibinfo}[2]{#2}
\providecommand{\eprint}[2][]{\url{#2}}

\bibitem[{\citenamefont{{M.A. Nielsen and I.L. Chuang}}(2000)}]{Nielsen:book}
\bibinfo{author}{\bibnamefont{{M.A. Nielsen and I.L. Chuang}}},
  \emph{\bibinfo{title}{{Quantum Computation and Quantum Information}}}
  (\bibinfo{publisher}{{Cambridge University Press}},
  \bibinfo{address}{Cambridge, UK}, \bibinfo{year}{2000}).

\bibitem[{\citenamefont{{D. Bacon, J. Kempe, D.A. Lidar and K.B.
  Whaley}}(2000)}]{Bacon:99aKempe:00}
\bibinfo{author}{\bibnamefont{{D. Bacon, J. Kempe, D.A. Lidar, and K.B.
  Whaley}}}, \bibinfo{journal}{Phys. Rev. Lett.} \textbf{\bibinfo{volume}{85}},
  \bibinfo{pages}{1758} (\bibinfo{year}{2000}); 
\bibinfo{author}{\bibnamefont{{J. Kempe, D. Bacon, D.A. Lidar, and K.B.
  Whaley}}}, \bibinfo{journal}{Phys. Rev. A} \textbf{\bibinfo{volume}{63}},
  \bibinfo{pages}{042307} (\bibinfo{year}{2001}).

\bibitem[{\citenamefont{{P. Zanardi and M. Rasetti}}(1997)}]{Zanardi:97cDuan:98Lidar:PRL98}
\bibinfo{author}{\bibnamefont{{P. Zanardi and M. Rasetti}}},
  \bibinfo{journal}{Phys. Rev. Lett.} \textbf{\bibinfo{volume}{79}},
  \bibinfo{pages}{3306} (\bibinfo{year}{1997});
\bibinfo{author}{\bibnamefont{{L.-M Duan and G.-C. Guo}}},
  \bibinfo{journal}{Phys. Rev. A} \textbf{\bibinfo{volume}{57}},
  \bibinfo{pages}{737} (\bibinfo{year}{1998});
\bibinfo{author}{\bibnamefont{{D.A. Lidar, I.L. Chuang, and K.B. Whaley}}},
  \bibinfo{journal}{Phys. Rev. Lett.} \textbf{\bibinfo{volume}{81}},
  \bibinfo{pages}{2594} (\bibinfo{year}{1998}).

\bibitem[{\citenamefont{{E. Knill, R. Laflamme, and L. Viola}}(2000)}]{Knill:99a}
\bibinfo{author}{\bibnamefont{{E. Knill, R. Laflamme, and L. Viola}}},
  \bibinfo{journal}{Phys. Rev. Lett.} \textbf{\bibinfo{volume}{84}},
  \bibinfo{pages}{2525} (\bibinfo{year}{2000}).

\bibitem[{\citenamefont{{D.P. DiVincenzo, D. Bacon, J. Kempe, G. Burkard, and
  K.B. Whaley}}(2000)}]{DiVincenzo:00a}
\bibinfo{author}{\bibnamefont{{D.P. DiVincenzo, D. Bacon, J. Kempe, G. Burkard, and K.B. Whaley}}}, \bibinfo{journal}{Nature}
  \textbf{\bibinfo{volume}{408}}, \bibinfo{pages}{339} (\bibinfo{year}{2000}).

\bibitem[{\citenamefont{{X. Hu, R. de Sousa, and S. Das Sarma}}(2001)}]{Hu:01a}
\bibinfo{author}{\bibnamefont{{X. Hu, R. de Sousa, and S. Das Sarma}}},
  \bibinfo{journal}{Phys. Rev. Lett.} \textbf{\bibinfo{volume}{86}},
  \bibinfo{pages}{918} (\bibinfo{year}{2001}).

\bibitem[{\citenamefont{{D.A. Lidar and L.-A. Wu}}(2002)}]{LidarWu:01}
\bibinfo{author}{\bibnamefont{{D.A. Lidar and L.-A. Wu}}},
  \bibinfo{journal}{Phys. Rev. Lett.} \textbf{\bibinfo{volume}{88}},
  \bibinfo{pages}{017905} (\bibinfo{year}{2002});
\bibinfo{author}{\bibnamefont{{L.-A. Wu and D.A. Lidar}}},
  \bibinfo{journal}{J. Math. Phys.} \textbf{\bibinfo{volume}{43}},
  \bibinfo{pages}{4506} (\bibinfo{year}{2002})

\bibitem[{\citenamefont{{J. Levy}}()}]{Levy:01Benjamin:01}
\bibinfo{author}{\bibnamefont{{J. Levy}}}, \bibinfo{journal}{Phys. Rev. Lett.} \textbf{\bibinfo{volume}{89}},
  \bibinfo{pages}{147902} (\bibinfo{year}{2002});
\bibinfo{author}{\bibnamefont{{S.C. Benjamin}}}, \bibinfo{journal}{Phys. Rev.
  A} \textbf{\bibinfo{volume}{64}}, \bibinfo{pages}{054303}
  (\bibinfo{year}{2001}).

\bibitem[{\citenamefont{{D. Loss and D.P. DiVincenzo}}(1998)}]{Loss:98Kane:98}
\bibinfo{author}{\bibnamefont{{D. Loss and D.P. DiVincenzo}}},
  \bibinfo{journal}{Phys. Rev. A} \textbf{\bibinfo{volume}{57}},
  \bibinfo{pages}{120} (\bibinfo{year}{1998});
\bibinfo{author}{\bibnamefont{{B.E. Kane}}}, \bibinfo{journal}{Nature}
  \textbf{\bibinfo{volume}{393}}, \bibinfo{pages}{133}
(\bibinfo{year}{1998}); \bibinfo{author}{\bibnamefont{{J. Levy}}},
  \bibinfo{journal}{Phys. Rev. A} \textbf{\bibinfo{volume}{64}},
  \bibinfo{pages}{052306} (\bibinfo{year}{2001});

\bibitem[{\citenamefont{{R. Vrijen, E. Yablonovitch, K. Wang, H.W. Jiang, A.
  Balandin, V. Roychowdhury, T. Mor, and D. DiVincenzo}}(2000)}]{Vrijen:00}
\bibinfo{author}{\bibnamefont{{R. Vrijen, E. Yablonovitch, K. Wang, H.W. Jiang, A.
  Balandin, V. Roychowdhury, T. Mor, and D. DiVincenzo}}},
  \bibinfo{journal}{Phys. Rev. A} \textbf{\bibinfo{volume}{62}},
  \bibinfo{pages}{012306} (\bibinfo{year}{2000}).

\bibitem[{\citenamefont{{K.V. Kavokin}}(2001)}]{Kavokin:01}
\bibinfo{author}{\bibnamefont{{K.V. Kavokin}}}, \bibinfo{journal}{Phys. Rev. B}
  \textbf{\bibinfo{volume}{64}}, \bibinfo{pages}{075305}
  (\bibinfo{year}{2001}).

\bibitem[{\citenamefont{{N.E. Bonesteel, D. Stepanenko, and D.P.
  DiVincenzo}}(2001)}]{Bonesteel:01}
\bibinfo{author}{\bibnamefont{{N.E. Bonesteel, D. Stepanenko, and D.P.
  DiVincenzo}}}, \bibinfo{journal}{Phys. Rev. Lett.}
  \textbf{\bibinfo{volume}{87}}, \bibinfo{pages}{207901}
(\bibinfo{year}{2001}).

\bibitem[{\citenamefont{{G. Burkard and D. Loss}}(2002)}]{Burkard:01}
\bibinfo{author}{\bibnamefont{{G. Burkard and D. Loss}}},
  \bibinfo{journal}{Phys. Rev. Lett.} \textbf{\bibinfo{volume}{88}},
  \bibinfo{pages}{047903} (\bibinfo{year}{2002}).

\bibitem[{\citenamefont{{D. Aharonov and M. Ben-Or}}()}]{Aharonov:99}
\bibinfo{author}{\bibnamefont{{D. Aharonov and M. Ben-Or}}},
  \bibinfo{note}{eprint quant-ph/9906129};
\bibinfo{author}{\bibnamefont{{J. Preskill}}},
  \bibinfo{journal}{Proc. Roy. Soc. London Ser. A} \textbf{\bibinfo{volume}{454}},
  \bibinfo{pages}{385} (\bibinfo{year}{1998});
\bibinfo{author}{\bibnamefont{{A.M. Steane}}},
  \bibinfo{note}{eprint quant-ph/0207119}.

\bibitem[{\citenamefont{{L.-A. Wu and
D.A. Lidar}}({\natexlab{a}})}]{WuLidar:01} \bibinfo{author}{\bibnamefont{{L.-A. Wu and D.A. Lidar}}},
  \bibinfo{journal}{Phys. Rev. A} \textbf{\bibinfo{volume}{65}},
  \bibinfo{pages}{042318} (\bibinfo{year}{2002}).

\bibitem[{WuL({\natexlab{a}})}]{comment2}
\bibinfo{note}{In deriving this formula we neglected an overall phase factor
  $e^{i \phi (1+\gamma \beta^2)S_1^z S_2^z}$ and employed the identity $e^{-i
  \theta S^z} S^x e^{i \theta S^z} = S^x \cos\theta + S^y \sin\theta$, which
  we use repeatedly below.}

\bibitem[{WuL({\natexlab{b}})}]{comment1}
\bibinfo{note}{Here we used the identity $e^{i\eta (S^{z}\cos 2\epsilon
  +S^{y}\sin 2\epsilon )} \times e^{-i\eta (S^{z}\cos 2\epsilon -S^{y}\sin 2\epsilon )}
= e^{i\xi (S^{x}\cos \delta +S^{y}\sin \delta )}$,
with angles as defined in the text.}

\bibitem[{\citenamefont{{J.L. Dodd, M.A. Nielsen, M.J. Bremner, and R.T. Thew}}()}]{Dodd:01}
\bibinfo{author}{\bibnamefont{{J.L. Dodd, M.A. Nielsen, M.J. Bremner, and R.T. Thew}}},  \bibinfo{journal}{Phys. Rev. A} \textbf{\bibinfo{volume}{65}},
\bibinfo{pages}{040301} (\bibinfo{year}{2002}).

\bibitem[{\citenamefont{{N. Khaneja, S.J. Glasser, and R. Brockett}}()}]{Khaneja:02}
\bibinfo{author}{\bibnamefont{{N. Khaneja, S.J. Glasser, and R. Brockett}}},  \bibinfo{journal}{Phys. Rev. A} \textbf{\bibinfo{volume}{65}},
\bibinfo{pages}{032301} (\bibinfo{year}{2002}).

\bibitem[{\citenamefont{{S. Lloyd}}(1995)}]{Lloyd:95}
\bibinfo{author}{\bibnamefont{{S. Lloyd}}}, \bibinfo{journal}{Phys. Rev. Lett.}
  \textbf{\bibinfo{volume}{75}}, \bibinfo{pages}{346} (\bibinfo{year}{1995}).

\bibitem[{\citenamefont{Viola and Lloyd}(1998)}]{Viola:98Vitali:99}
\bibinfo{author}{\bibfnamefont{L.}~\bibnamefont{Viola}} \bibnamefont{and}
  \bibinfo{author}{\bibfnamefont{S.}~\bibnamefont{Lloyd}},
  \bibinfo{journal}{Phys. Rev. A} \textbf{\bibinfo{volume}{58}},
\bibinfo{pages}{2733} (\bibinfo{year}{1998}); \bibinfo{author}{\bibfnamefont{D.}~\bibnamefont{Vitali}} \bibnamefont{and}
  \bibinfo{author}{\bibfnamefont{P.}~\bibnamefont{Tombesi}},
  \bibinfo{journal}{Phys. Rev. A} \textbf{\bibinfo{volume}{59}},
  \bibinfo{pages}{4178} (\bibinfo{year}{1999});
\bibinfo{author}{\bibnamefont{{P. Zanardi}}},
\bibinfo{journal}{Phys. Lett. A}
  \textbf{\bibinfo{volume}{258}}, \bibinfo{pages}{77}
(\bibinfo{year}{1999}); \bibinfo{author}{\bibnamefont{{M.S. Byrd and D.A. Lidar}}}, \bibinfo{journal}{Quant. Inf. Proc.}
  \textbf{\bibinfo{volume}{1}}, \bibinfo{pages}{19} (\bibinfo{year}{2002}).

\bibitem[{\citenamefont{{L.-A. Wu and D.A.
  Lidar}}({\natexlab{b}})}]{WuLidar:01b}
\bibinfo{author}{\bibnamefont{{L.-A. Wu and D.A. Lidar}}},
  \bibinfo{journal}{Phys. Rev. Lett.} \textbf{\bibinfo{volume}{88}},
\bibinfo{pages}{207902} (\bibinfo{year}{2002}).

\bibitem[{\citenamefont{{J. Kempe, D. Bacon, D.P. DiVincenzo, and K.B.
  Whaley}}()}]{Kempe:01WuByrdLidar:02}
\bibinfo{author}{\bibnamefont{{J. Kempe, D. Bacon, D.P. DiVincenzo, and K.B.
  Whaley}}},
\bibinfo{journal}{Quant. Inf. Comp.}\textbf{\bibinfo{volume}{1}},
\bibinfo{pages}{33} (\bibinfo{year}{2001});
\bibinfo{author}{\bibnamefont{{M.S. Byrd and D.A. Lidar}}}, \bibinfo{journal}{Phys. Rev. Lett.} \textbf{\bibinfo{volume}{89}},
\bibinfo{pages}{047901} (\bibinfo{year}{2002}); 
\bibinfo{author}{\bibnamefont{{L.-A. Wu, M.S. Byrd, and D.A. Lidar}}}, \bibinfo{journal}{Phys. Rev. Lett.} \textbf{\bibinfo{volume}{89}},
\bibinfo{pages}{127901} (\bibinfo{year}{2002}).

\end{thebibliography}

\end{document}